\documentclass[journal]{IEEEtran}

\usepackage{graphicx, subcaption}
\graphicspath{{figs/}} 

\usepackage[nowarn,acronyms,nonumberlist,nopostdot,nomain,nogroupskip]{glossaries}
\usepackage{multirow}
\usepackage{setspace}
\usepackage{amssymb}
\usepackage{mathtools}

\DeclarePairedDelimiter\floor{\lfloor}{\rfloor}
\usepackage{algorithm}
\usepackage{algpseudocode}
\usepackage{tikz}

\newacronym{3gpp}{3GPP}{3rd Generation Partnership Project}
\newacronym{5g}{5G}{5\textsuperscript{th} Generation}
\newacronym{5gc}{5GC}{5G Core}
\newacronym{bs}{BS}{Base Station}
\newacronym{abft}{A-BFT}{Association-BeamForming Training}
\newacronym[firstplural=Access Categories (ACs)]{ac}{AC}{Access Category}
\newacronym{adc}{ADC}{Analog to Digital Converter}
\newacronym{addts}{ADDTS}{Add Traffic Stream}
\newacronym{afbw}{AFBW}{Average Fading Bandwidth}
\newacronym{aid}{AID}{Association Identifier}
\newacronym{aimd}{AIMD}{Additive Increase Multiplicative Decrease}
\newacronym{am}{AM}{Acknowledged Mode}
\newacronym{amc}{AMC}{Adaptive Modulation and Coding}
\newacronym{ampdu}{A-MPDU}{MAC Protocol Data Unit Aggregation}
\newacronym{mmse}{MMSE}{Minimum Mean Square Error}
\newacronym{aoa}{AoA}{Angle of Arrival}
\newacronym{aod}{AoD}{Angle of Departure}
\newacronym{ap}{AP}{Access Point}
\newacronym{app}{APP}{Application Layer}
\newacronym{aqm}{AQM}{Active Queue Management}
\newacronym{ar}{AR}{Augmented Reality}
\newacronym{ati}{ATI}{Announcement Transmission Interval}
\newacronym{awgn}{AGWN}{Additive White Gaussian Noise}
\newacronym{awv}{AWV}{Antenna Weight Vector}
\newacronym{balia}{BALIA}{Balanced Link Adaptation}
\newacronym{bdp}{BDP}{Bandwidth-Delay Product}
\newacronym{bf}{BF}{Beamforming}
\newacronym{bhi}{BHI}{Beacon Header Interval}
\newacronym{bi}{BI}{Beacon Interval}
\newacronym{brp}{BRP}{Beam Refinement Protocol}
\newacronym{bss}{BSS}{Basic Service Set}
\newacronym{bti}{BTI}{Beacon Transmission Interval}
\newacronym{cad}{CAD}{Computer-aided Design}
\newacronym{cbap}{CBAP}{Contention-Based Access Period}
\newacronym{cbr}{CBR}{Constant Bitrate}
\newacronym{cc}{CC}{Congestion Control}
\newacronym{cdf}{CDF}{Cumulative Distribution Function}
\newacronym{cf}{CF}{Cell-free}
\newacronym{cfmmimo}{CF mMIMO}{\gls{cf} \gls{mmimo}}
\newacronym{cir}{CIR}{Channel Impulse Response}
\newacronym{cn}{CN}{Core Network}
\newacronym{cp}{CP}{Control Plane}
\newacronym{cpu}{CPU}{central processing units}
\newacronym{cqi}{CQI}{Channel Quality Indicator}
\newacronym{crs}{CRS}{Cell Reference Signal}
\newacronym{csirs}{CSI-RS}{Channel State Information - Reference Signal}
\newacronym{csmaca}{CSMA/CA}{Carrier Sense Multiple Access with Collision Avoidance}
\newacronym{cts}{CTS}{Clear to Send}
\newacronym{d2d}{D2D}{Device-to-device}
\newacronym{dc}{DC}{Dual Connectivity}
\newacronym{dce}{DCE}{Direct Code Execution}
\newacronym{dcf}{DCF}{Distributed Coordination Function}
\newacronym{dci}{DCI}{Downlink Control Information}
\newacronym{delts}{DELTS}{Delete Traffic Stream}
\newacronym{dked}{DKED}{Double Knife Edge Diffraction}
\newacronym{dl}{DL}{Downlink}
\newacronym{dmg}{DMG}{Directional Multi-Gigabit}
\newacronym{dmr}{DMR}{Deadline Miss Ratio}
\newacronym{dmrs}{DMRS}{DeModulation Reference Signal}
\newacronym{dpp}{DPP}{Determinantal Point Processes}
\newacronym{dti}{DTI}{Data Transmission Interval}
\newacronym{dtmke}{DTMKE}{Double-truncated Multiple Knife-edge}
\newacronym{e2e}{E2E}{End-to-End}
\newacronym{ecn}{ECN}{Explicit Congestion Notification}
\newacronym{edca}{EDCA}{Enhanced Distributed Channel Access}
\newacronym{edf}{EDF}{Earliest Deadline First}
\newacronym{enb}{eNB}{evolved Node Base}
\newacronym{endc}{EN-DC}{E-UTRAN-\gls{nr} \gls{dc}}
\newacronym{epc}{EPC}{Evolved Packet Core}
\newacronym{es}{ES}{Edge Server}
\newacronym{ese}{ESE}{Extended Schedule Element}
\newacronym{fdd}{FDD}{Frequency Division Duplexing}
\newacronym{fdma}{FDMA}{Frequency Division Multiple Access}
\newacronym{fov}{FoV}{Field-of-View}
\newacronym{fs}{FS}{Fast Switching}
\newacronym{ftp}{FTP}{File Transfer Protocol}
\newacronym{gnb}{gNB}{Next Generation Node Base}
\newacronym{harq}{HARQ}{Hybrid Automatic Repeat reQuest}
\newacronym{hetnet}{HetNet}{Heterogeneous Network}
\newacronym{hh}{HH}{Hard Handover}
\newacronym{hol}{HOL}{Head-of-Line}
\newacronym{hqf}{HQF}{Highest-quality-first}
\newacronym{ia}{IA}{Initial Access}
\newacronym{iab}{IAB}{Integrated Access and Backhaul}
\newacronym{ibss}{IBSS}{Independent Basic Service Set}
\newacronym{id}{ID}{Identifier}
\newacronym{imt}{IMT}{International Mobile Telecommunication}
\newacronym{inr}{INR}{Interference to Noise Ratio}
\newacronym{iot}{IoT}{Internet of Things}
\newacronym{ipa}{IPA}{Inter-Packet Arrival}
\newacronym{ism}{ISM}{Industrial, Scientific, and Medical}
\newacronym{kpi}{KPI}{Key Performance Indicator}
\newacronym{lcf}{LCF}{Level Crossing Frequency}
\newacronym{lcr}{LCR}{Level Crossing Rate}
\newacronym{los}{LoS}{Line-of-Sight}
\newacronym{lp}{LP}{Low Power}
\newacronym{lsf}{LSF}{large-scale fading}
\newacronym{lte}{LTE}{Long Term Evolution}
\newacronym{m2m}{M2M}{Machine to Machine}
\newacronym{mac}{MAC}{Medium Access Control}
\newacronym{mc}{MC}{Multi-Connectivity}
\newacronym{mcs}{MCS}{Modulation and Coding Scheme}
\newacronym{mec}{MEC}{Mobile Edge Cloud}
\newacronym{mi}{MI}{Mutual Information}
\newacronym{mib}{MIB}{Master Information Block}
\newacronym{mimo}{MIMO}{Multiple Input, Multiple Output}
\newacronym{mmimo}{mMIMO}{massive Multiple-Input, Multiple-Output}
\newacronym{mumimo}{MU-MIMO}{Multi-User Multiple Input, Multiple Output}
\newacronym{ml}{ML}{Machine Learning}
\newacronym{mlr}{MLR}{Maximum-local-rate}
\newacronym[plural=\gls{mme}s,firstplural=Mobility Management Entities (MMEs)]{mme}{MME}{Mobility Management Entity}
\newacronym{mmwave}{mmWave}{Millimeter Wave}
\newacronym{moi}{MoI}{Method of Images}
\newacronym{mpc}{MPC}{Multi Path Component}
\newacronym{mptcp}{MPTCP}{Multipath TCP}
\newacronym{mr}{MR}{Maximum Ratio}
\newacronym{mrdc}{MR-DC}{Multi \gls{rat} \gls{dc}}
\newacronym{mss}{MSS}{Maximum Segment Size}
\newacronym{mtd}{MTD}{Machine-Type Device}
\newacronym{mtu}{MTU}{Maximum Transmission Unit}
\newacronym{nav}{NAV}{Network Allocation Vector}
\newacronym{ncbr}{NCBR}{Non-Constant Bitrate}
\newacronym{nfv}{NFV}{Network Function Virtualization}
\newacronym{nlos}{NLoS}{Non-Line-of-Sight}
\newacronym{nr}{NR}{New Radio}
\newacronym{nrmse}{NRMSE}{Normalized Root Mean Square Error}
\newacronym{ns3}{ns-3}{Network Simulator 3}
\newacronym{nsa}{NSA}{Non Stand Alone}
\newacronym{o2i}{O2I}{Outdoor-to-Indoor}
\newacronym{ofdm}{OFDM}{Orthogonal Frequency Division Multiplexing}
\newacronym{pa}{PA}{Position-aware}
\newacronym{pan}{PAN}{Personal Area Network}
\newacronym{pas}{PAS}{Power Angular Spectrum}
\newacronym{pbch}{PBCH}{Physical Broadcast Channel}
\newacronym{pbss}{PBSS}{Personal Basic Service Set}
\newacronym{pci}{PCI}{Physical Cell Identity}
\newacronym{pcp}{PCP}{\gls{pbss} Central Point}
\newacronym{pcpap}{PCP/AP}{\acrlong{pcp}/\acrlong{ap}}
\newacronym{pdcch}{PDCCH}{Physical Downlonk Control Channel}
\newacronym{pdcp}{PDCP}{Packet Data Convergence Protocol}
\newacronym{pdsch}{PDSCH}{Physical Downlink Shared Channel}
\newacronym{pdu}{PDU}{Packet Data Unit}
\newacronym{pf}{PF}{Proportional Fair}
\newacronym{pgw}{PGW}{Packet Gateway}
\newacronym{phy}{PHY}{Physical Layer}
\newacronym{ppp}{PPP}{Poisson Point Process}
\newacronym{prb}{PRB}{Physical Resource Block}
\newacronym{pss}{PSS}{Primary Synchronization Signal}
\newacronym{pucch}{PUCCH}{Physical Uplink Control Channel}
\newacronym{pusch}{PUSCH}{Physical Uplink Shared Channel}
\newacronym{qd}{QD}{Quasi Deterministic}
\newacronym{qoe}{QoE}{Quality of Experience}
\newacronym{qos}{QoS}{Quality of Service}
\newacronym{rach}{RACH}{Random Access Channel}
\newacronym{ran}{RAN}{Radio Access Network}
\newacronym[firstplural=Radio Access Technologies (RATs)]{rat}{RAT}{Radio Access Technology}
\newacronym{red}{RED}{Random Early Detection}
\newacronym{rf}{RF}{Radio Frequency}
\newacronym{rl}{RL}{Reinforcement Learning}
\newacronym{rlc}{RLC}{Radio Link Control}
\newacronym{rlf}{RLF}{Radio Link Failure}
\newacronym{rms}{RMS}{Root Mean Square}
\newacronym{rr}{RR}{Round Robin}
\newacronym{rrc}{RRC}{Radio Resource Control}
\newacronym{rrm}{RRM}{Radio Resource Management}
\newacronym{rs}{RS}{Remote Server}
\newacronym{rsrp}{RSRP}{Reference Signal Received Power}
\newacronym{rsrq}{RSRQ}{Reference Signal Received Quality}
\newacronym{rss}{RSS}{Received Signal Strength}
\newacronym{rssi}{RSSI}{Received Signal Strength Indicator}
\newacronym{rt}{RT}{Ray Tracer}
\newacronym{rts}{RTS}{Request to Send}
\newacronym{rtt}{RTT}{Round Trip Time}
\newacronym{rw}{RW}{Receive Window}
\newacronym{rx}{RX}{Receiver}
\newacronym{sa}{SA}{standalone}
\newacronym{sack}{SACK}{Selective Acknowledgment}
\newacronym{sap}{SAP}{Service Access Point}
\newacronym{sc}{SC}{Single Carrier}
\newacronym{sch}{SCH}{Secondary Cell Handover}
\newacronym{scm}{SCM}{Spatial Channel Model}
\newacronym{scoot}{SCOOT}{Split Cycle Offset Optimization Technique}
\newacronym{sdma}{SDMA}{Spatial Division Multiple Access}
\newacronym{sdr}{SDR}{Software Defined Radio}
\newacronym{se}{SE}{Spectral Efficiency}
\newacronym{si}{SI}{Study Item}
\newacronym{sib}{SIB}{Secondary Information Block}
\newacronym{sinr}{SINR}{Signal-to-Interference-plus-Noise Ratio}
\newacronym{sir}{SIR}{Signal-to-Interference Ratio}
\newacronym{sls}{SLS}{Sector-Level Sweep}
\newacronym{sm}{SM}{Saturation Mode}
\newacronym{snr}{SNR}{Signal-to-Noise Ratio}
\newacronym{son}{SON}{Self-Organizing Network}
\newacronym{sp}{SP}{Service Period}
\newacronym{spr}{SPR}{Service Period Request}
\newacronym{srs}{SRS}{Sounding Reference Signal}
\newacronym{ss}{SS}{Synchronization Signal}
\newacronym{ssb}{SSB}{\gls{ss}}
\newacronym{sss}{SSS}{Secondary Synchronization Signal}
\newacronym{ssw}{SSW}{Sector Sweep}
\newacronym{sta}{STA}{Station}
\newacronym{stb}{STB}{Set Top Box}
\newacronym{tb}{TB}{Transport Block}
\newacronym{tcp}{TCP}{Transmission Control Protocol}
\newacronym{tdd}{TDD}{Time Division Duplexing}
\newacronym{tdma}{TDMA}{Time Division Multiple Access}
\newacronym{tfl}{TfL}{Transport for London}
\newacronym{tgad}{TGad}{Task Group ad}
\newacronym{tgay}{TGay}{Task Group ay}
\newacronym{tsconst}{TSCONST}{Traffic Scheduling Constraint}
\newacronym{tm}{TM}{Transparent Mode}
\newacronym{trp}{TRP}{Transmitter Receiver Pair}
\newacronym{ts}{TS}{Traffic Stream}
\newacronym{tspec}{TSPEC}{Traffic Specification}
\newacronym{tti}{TTI}{Transmission Time Interval}
\newacronym{ttt}{TTT}{Time-to-Trigger}
\newacronym{tx}{TX}{Transmitter}
\newacronym[firstplural=Transmission Opportunities (TXOPs)]{txop}{TXOP}{Transmission Opportunity}
\newacronym{udp}{UDP}{User Datagram Protocol}
\newacronym{ue}{UE}{User Equipment}
\newacronym{ul}{UL}{Uplink}
\newacronym{um}{UM}{Unacknowledged Mode}
\newacronym{uma}{UMa}{Urban Macro}
\newacronym{uml}{UML}{Unified Modeling Language}
\newacronym{utc}{UTC}{Urban Traffic Control}
\newacronym{v2v}{V2V}{Vehicle-to-Vehicle}
\newacronym{vbr}{VBR}{Variable Bit Rate}
\newacronym{vm}{VM}{Virtual Machine}
\newacronym{vr}{VR}{Virtual Reality}
\newacronym{wbf}{WBF}{Wired Bias Function}
\newacronym{wf}{WF}{Wired-first}
\newacronym{wifi}{Wi-Fi}{Wireless Fidelity}
\newacronym{wigig}{WiGig}{Wireless Gigabit}
\newacronym{wlan}{WLAN}{Wireless Local Area Network}
\newacronym{ber}{BER}{Bit Error Rate}
\newacronym{arf}{ARF}{Auto Rate Fallback}
\newacronym{semm}{SEMM}{SPCA-EDCA Mixed Mode}
\newacronym{ppdu}{PPDU}{PHY Protocol Data Unit}
\newacronym{udn}{UDN}{Ultra Dense Network}
\newacronym{dnn}{DNN}{Deep Neural Network}

\usepackage{xcolor}

\usepackage{bbold}

\usepackage{cleveref}
\Crefformat{figure}{#2Fig.~#1#3}
\Crefformat{table}{#2Tab.~#1#3}
\Crefformat{section}{#2Sec.~#1#3}
\Crefmultiformat{figure}{Figs.~#2#1#3}{ and~#2#1#3}{, #2#1#3}{ and~#2#1#3}

\makeglossaries
\glsdisablehyper
\usepackage{cite}

\begin{document}

\title{Repulsive Clustering Based Pilot Assignment for Cell-Free Massive MIMO Systems
\thanks{Salman Mohebi has received funding from the European Union’s Horizon 2020 research and innovation programme under the Marie Skłodowska-Curie Grant agreement No. 813999.}}
\author{Salman Mohebi, Andrea Zanella, Michele Zorzi\\
Department of Information Engineering, University of Padova, Padova, Italy \\
E-mails: \texttt{\{surname\}@dei.unipd.it}}

\maketitle
\pagenumbering{gobble}


\begin{abstract}
Thanks to its capability to provide a uniform service rate for the \glspl{ue}, \gls{cf} \gls{mmimo}, has recently attracted considerable attention, both in academia and in industry, and so is considered as one of the potential technologies for beyond-5G and 6G.
However, the reuse of the same pilot signals by multiple users can create the so-called pilot contamination problem, which can hinder the \gls{cf} \gls{mmimo} from unlocking its full performance.
In this paper, we address the challenge by formulating the pilot assignment as a maximally diverse clustering problem and propose an efficient yet straightforward repulsive clustering-based pilot assignment scheme to mitigate the effects of pilot contamination
on \gls{cf} \gls{mmimo}.
The numerical results show the superiority of the proposed technique compared to some other methods with respect to the achieved uplink per-user rate.
\end{abstract}

\begin{IEEEkeywords}
cell-free massive MIMO, pilot assignment, pilot contamination, repulsive clustering, maximally diverse clustering
\end{IEEEkeywords}

\IEEEpeerreviewmaketitle

\begin{tikzpicture}[remember picture,overlay]
\node[anchor=north,yshift=-10pt] at (current page.north) {\parbox{\dimexpr\textwidth-\fboxsep-\fboxrule\relax}{
\centering\footnotesize This work has been submitted to the EUSIPCO 2022 for possible publication. Copyright may be
transferred without notice.}};
\end{tikzpicture}

\section{Introduction}\label{sec:intro}
\glsresetall
Evolving from the first to the fifth generation, many technologies have been proposed to support growing traffic and service demands in mobile networks.
Network densification is a common technique to increase the network coverage and rate for the \glspl{ue}.
Densification can happen both by increasing the number of the \glspl{bs}, a.k.a. ultra dense networks, or the number of the antennas at the \gls{bs}, a.k.a. \gls{mmimo}.
Each of these approaches suffers from some shortages:
deploying a large number of \glspl{bs} increases the inter-cell interference and hence reduces the service quality for the \glspl{ue}, while in the \gls{mmimo}, \glspl{ue}  located at the edge of the cell suffer from high propagation loss because of the long distance from the \gls{bs}.
\gls{cf} \gls{mmimo}~\cite{ngo2017cell} has recently been introduced as an answer to the shortage of the technologies mentioned above by adopting the best of both.
The \gls{cf} \gls{mmimo} systems are composed of a large number of distributed \glspl{ap} that jointly serve relatively fewer number of \glspl{ue}.
The operation, unlike the traditional cellular network, takes place in a user-centric fashion, where each \gls{ue} is surrounded and served by multiple \glspl{ap}.
The \glspl{ap} are connected to a \gls{cpu} through high-capacity error-free channels, where the network synchronization, data detection/precoding/decoding, and some other network management operations take place.

\gls{cf} \gls{mmimo} adopts the block fading models, where time-frequency channels are divided into coherence blocks of $\tau_c$ channel uses.
Each coherent block is further divided into three sub-intervals such that: $\tau_c = \tau_p + \tau_u + \tau_d$, where $\tau_p$ is used for uplink pilot training, and $\tau_u$ and $\tau_d$ are used for uplink and downlink data transmission, respectively.
Due to the limited number of channel uses in each coherence block, we can only have a limited number of orthogonal pilots, which is typically smaller than the number of \glspl{ue}. 
This forces us to reuse the same pilots for different \glspl{ue}, which introduces some undesirable effects, known as pilot contamination: the fading channel can not be accurately estimated at the \glspl{ap} due to the co-pilot interference among \glspl{ue}.

The \textit{random} pilot assignment presented in~\cite{ngo2017cell} is not efficient, and a proper pilot assignment policy can significantly reduce the effects of the so-called pilot contamination problem.
A \textit{greedy} pilot assignment is proposed in~\cite{ngo2017cell}, which iteratively updates the pilot sequence for the \gls{ue} with minimum rate.
A \textit{structured} pilot assignment scheme is proposed in~\cite{attarifar2018random} that maximizes the minimum distance between the co-pilot \glspl{ue}.
A \textit{location-based greedy} pilot assignment is proposed in~\cite{zhang2018location}, that utilized the location information of the \glspl{ue} to improve the initial pilot assignment.
The authors in~\cite{yu2022topological} considered the pilot assignment as a topological interference management problem with multiple groupcasting messages. They then formulated two \textit{topological} pilot assignments for known and unknown \gls{ue}/\gls{ap} connectivity patterns.
Graph theory has also been used for modeling the pilot assignment, where by creating interference graph among the \glspl{ue}, \textit{graph coloring},~\cite{liu2020graph} and \textit{weight graphic} \cite{zeng2021pilot} is used to assign pilots for different \glspl{ue}.
\textit{Tabu search} is another approach that has already been considered to the pilot assignment problem~\cite{liu2019tabu}.
Buzzi~\emph{et.~al.}~\cite{buzzi2020pilot} formulated pilot assignment as a graph matching problem and proposed a \textit{Hungarian} algorithm to solve it.
A weighted count-based pilot assignment is presented in~\cite{li2021pilot}, which considers the user's prior geographic information and pilot power to maximize the pilot reuse weighted distance.
Another scalable pilot assignment algorithm based on \textit{deep learning} is presented in~\cite{li2021scalable} that maps between user locations and pilot assignment schemes.
The co-pilot interference, in principle, is because of the pilot reuse in \glspl{ue} that are close to each other.
So, a valid pilot assignment scheme could only rely on the \glspl{ue} geographical locations instead of adopting a costly channel estimation procedure to form the interference graph.
Motivated by the above considerations, in this paper, we consider the pilot assignment in \gls{cf} \gls{mmimo} as a maximally diverse clustering problem, where the \glspl{ue} are divided into clusters that maximize inter-cluster heterogeneity and intra-cluster homogeneity. 
We then present a repulsive clustering method to solve it.
The remainder of this paper is summarized as follows.
\Cref{sec:sysmodel} provides the system model for the \gls{cf} \gls{mmimo}, then in \Cref{sec:problem_formulation}, we formulate the pilot assignment problem and propose the repulsive clustering based pilot assignment.
The numerical results are presented in \Cref{sec:result}, and we conclude the paper in \Cref{sec:conclusion}.



\section{System Model}\label{sec:sysmodel}
\begin{figure}[t]
    \centering
    \includegraphics[width=\linewidth]{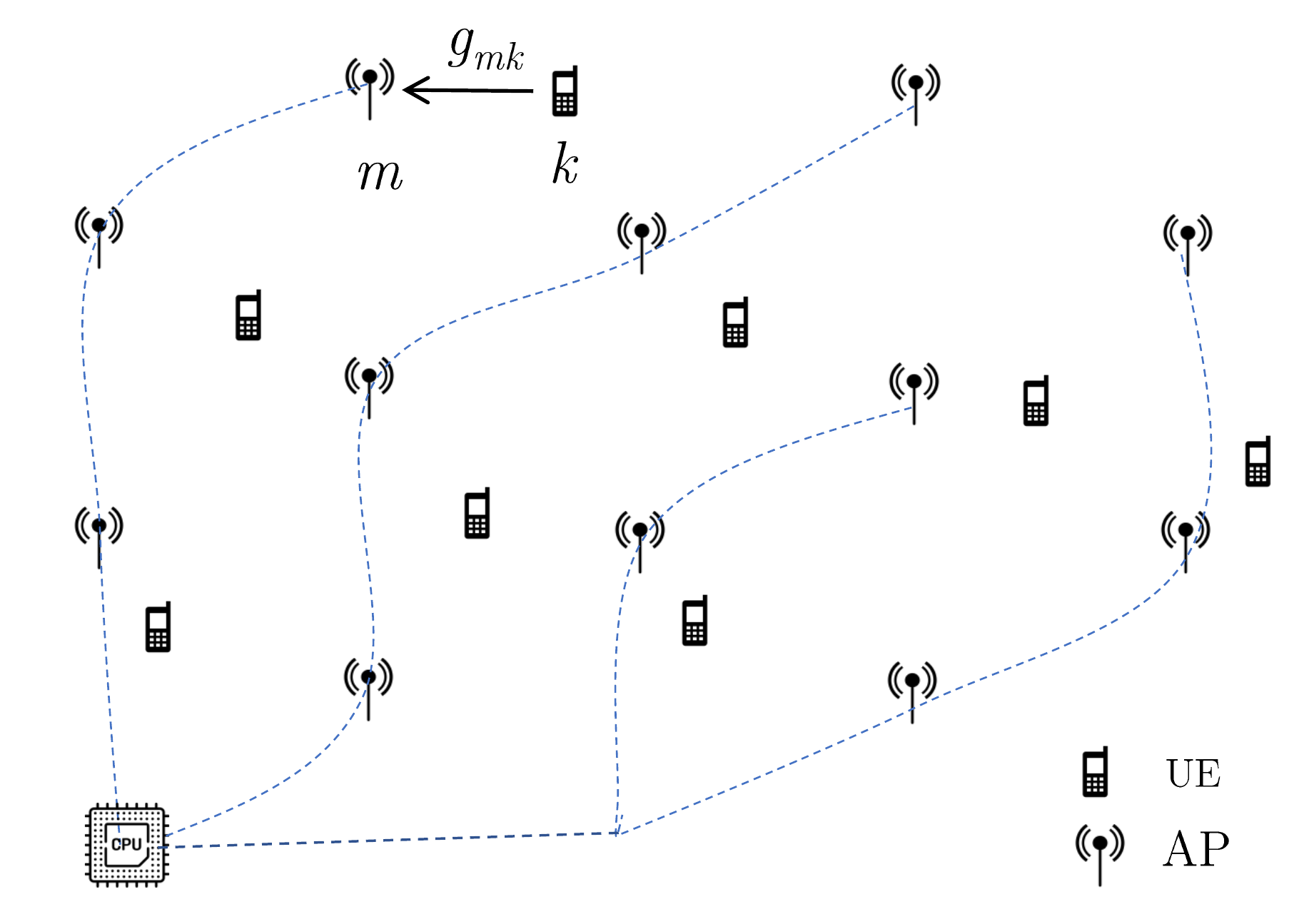}
    \caption{A CF mMIMO system, where $M$ distributed APs jointly serve $K$ UEs ($K<<M$).}
    \label{fig:system_model}
\end{figure}

We consider a typical \gls{cf} \gls{mmimo} system, where $M$ geographically distributed \glspl{ap} equipped with single antenna, coherently serve $K$ single-antenna \gls{ue} ($K << M$),  as exemplified in~\Cref{fig:system_model}.
All \glspl{ap} are connected to a \gls{cpu} by an unlimited error-free fronthaul channel.
The channel coefficient $g_{mk}$  between the $m$-th \gls{ap} and the $k$-th \gls{ue} is given as follows:
\begin{equation}
g_{mk} = \beta_{mk}^{1/2}h_{mk},
\end{equation}
where $\{\beta_{mk}\}$ indicate the \gls{lsf} coefficients (i.e., pathloss and shadowing), and $\{h_{mk}\}$ represent the small-scale fading coefficient which are assumed to be independent identically distributed (i.i.d.) normal random variables $\mathcal{CN}(0,1)$.

\subsection{Uplink Pilot Training}
We assume that there are only $\tau_p$ mutually orthogonal pilot sequences with length $\tau_p$ each represented as a column $\phi \in C^{\tau_p \times 1}$ of a matrix $\mathbf{\Phi}$, for which we have $\Vert\phi_{p_k}^{H}\phi_{p_{k'}}\Vert = 1$ if $p_k=p_{k'}$, and $\Vert\phi_{p_k}^{H}\phi_{p_{k'}}\Vert = 0$, otherwise.
The number of available pilots is independent of $K$ and is limited due to the natural channel variation in the time and frequency domains~\cite{chen2021survey}.

In the uplink pilot training phase, all \glspl{ue} simultaneously transmit their pilots.
The $m$-th \gls{ap} receives

\begin{equation}
{y}_m^p=\sqrt{\tau_{p} \rho_{p}} \sum_{k=1}^{K} g_{m k} \boldsymbol{\phi}_{p_{k}}^{H}+n_m^p, 
\end{equation}
where $\rho_p$ is the normalized \gls{snr} of a pilot sequence with respect to noise power, and $n_m^p \sim\mathcal{CN}(0, 1)$ represents the additive thermal noise. 

As shown in~\cite{ngo2017cell}, the effective channel coefficients between \gls{ue} $k$ and \gls{ap} $m$ can be estimated employing \gls{mmse} estimator as follows:

\begin{equation}
\hat{g}_{m k}\!=\!c_{m k}\!\Big(\!\sqrt{\tau_{p} \rho_{p}} g_{m k}\!+\!\sqrt{\tau_{p} \rho_{p}} \!\sum_{k^{\prime} \neq k}^{K} g_{m k^{\prime}} \boldsymbol{\phi}_{p_{k}}^{H} \boldsymbol{\phi}_{p_{k^{\prime}}}\!+\!\boldsymbol{\phi}_{p_{k}}\! n_m^p\!\Big)\!,
\end{equation}
where
\begin{equation}
c_{m k} \triangleq \frac{\sqrt{\tau_{p} \rho_{p}} \beta_{m k}}{\tau_{p} \rho_{p} \sum_{k^{\prime}=1}^{K} \beta_{m k^{\prime}}\left|\phi_{p_{k}}^{H} \phi_{p_{k^{\prime}}}\right|^{2}+1}.
\end{equation}

The mean-square of the estimated channel vector $\hat{g}_{mk}$ is
\begin{equation}
\gamma_{m k} \triangleq \mathbb{E}\left\{\left|\hat{g}_{m k}\right|^{2}\right\}=\sqrt{\tau_{p} \rho_{p}} \beta_{m k} c_{m k}.
\end{equation}

\subsection{Uplink Data Transmission}
In \gls{cf} \gls{mmimo}, all \glspl{ap} and \glspl{ue} use the same time-frequency resources to transmit data.
In the uplink, \gls{ap} $m$ receives
\begin{equation}
{y}_m^u=\sqrt{\rho_u} \sum_{k=1}^{K} g_{m k} \sqrt{\eta_k}q_k+n_m^u, 
\end{equation}
where $q_k$ is the signal transmitted by \gls{ue} $k$ with power $\mathbb{E}\left\{\left|q_k\right|^{2}\right\}=1$, while $\eta\in\left[0,1\right]$ indicates the power control coefficient, $\rho_u$ denotes the normalized uplink \gls{snr} and $n_m^u \sim\mathcal{CN}(0, 1)$ is the additive noise at receiver.

The \gls{mr} combining scheme can be applied to decode the desired signal from a certain \gls{ue} $k$.
\gls{ap} $m$ sends $\hat{g}^*_{mk}y^u_m$ to the \gls{cpu} for data detection.
The \gls{cpu} combines all the received signal for \gls{ue} $k$ as: $r^u_k = \sum_{m=1}^M\hat{g}^*_{mk}y^u_m$.
The signal then can be decomposed at the \gls{cpu} as follows, as shown in \cite{ngo2017cell}:
\begin{equation}
\begin{aligned}
r_k^u=& \underbrace{\sqrt{\rho_{u} \eta_{k}} q_{k} \mathbb{E}\left\{\sum_{m=1}^{M} g_{m k} \hat{g}_{m k}^{*}\right\}}_{\mathrm{DS}_{k}}\\
&+\underbrace{\sqrt{\rho_{u} \eta_{k}} q_{k}\left(\sum_{m=1}^{M} g_{m k} \hat{g}_{m k}^{*}-\mathbb{E}\left\{\sum_{m=1}^{M} g_{m k} \hat{g}_{m k}^{*}\right\}\right)}_{\mathrm{BU}_{k}} \\
&+\underbrace{\sqrt{\rho_{u}} \sum_{m=1}^{M} \sum_{k^{\prime} \neq k}^{K} \sqrt{\eta_{k^{\prime}}} g_{m k} \hat{g}_{m k^{\prime}}^{*} q_{k^{\prime}}}_{\mathrm{CPI}_{k}}+\hat{g}_{m k}^{*} n_k^u,
\end{aligned}
\end{equation}
where $\text{DS}_k$, $\text{BU}_k$ and $\text{CPI}_k$ denoted the desired signal (DS), beamforming uncertainty (BU) and co-pilot interference (CPI), respectively.

The achievable uplink rate for the \gls{ue} $k$ can be calculated as \eqref{eq:ul_sinr}, shown at the top of the next page.
\begin{table*}
\begin{align}\label{eq:ul_sinr}
    \operatorname{R}_k^u=\log_2\left(1+\frac{\rho_{u} \eta_{k}\left(\sum_{m=1}^M \gamma_{m k}\right)^{2}}{\rho_{u} \sum_{k^{\prime} \neq k}^{K} \eta_{k^{\prime}}\left(\sum_{m=1}^{M} \gamma_{m k} \frac{\beta_{m k^{\prime}}}{\beta_{m k}}\right)^2\left|\phi_{p_{k}}^{H} \phi_{p_{k^{\prime}}}\right|+\rho_{u} \sum_{k^{\prime}=1}^{K} \eta_{k^{\prime}} \sum_{m=1}^{M} \gamma_{m k} \beta_{m k^{\prime}}+\sum_{m=1}^{M} \gamma_{m k}}\right)
\end{align}
\hrule
\end{table*}

\section{Pilot-Assignment and Serving Cluster Formation}\label{sec:problem_formulation}
\subsection{Problem formulation}

An efficient pilot assignment mechanism should maximize the number of effectively estimated channels between the \glspl{ue} and the \glspl{ap}.
Due to the coherent nature of the transmissions in \gls{cf} \gls{mmimo} systems, data can still be potentially detected in the presence of multiple imperfectly estimated channels.
So, as in \gls{cf} \gls{mmimo} the ultimate goal is increasing the rate for the \glspl{ue}, the pilot assignment can be formulated as an uplink rate maximization problem, i.e.,

\begin{equation}
\begin{aligned}
    \max_{\textbf{p}} \quad & \sum_{k=1}^K \operatorname{R}_k^u \\
    \textrm{s.t.} \quad & \textbf{p} = \{p_1, ...p_K\} \\
    \quad & \phi \in \mathbf{\Phi}, \quad \forall k\in\{1,\ldots,K\}.
\end{aligned}
\end{equation}

\subsection{Proposed Scheme}
Considering that the distance between \glspl{ue} has a significant impact on co-pilot interference, to mitigate the effects of pilot contamination an efficient pilot assignment policy should assign the same pilot $p$ to \glspl{ue} in a repulsive way, i.e., to the \glspl{ue} that are geographically far apart or have fewer common serving \glspl{ap}.
Hence, we formulate the pilot assignment as a maximally diverse clustering problem, where the data points (\glspl{ue}) that are assigned to the same cluster have high "dissimilarity", but can be similar to the members from different clusters.
To solve the problem, we then proposed a repulsive clustering scheme, that is opposed to typical clustering algorithms which put homogeneous data points in the same clusters.
Note that, the inter-cluster similarity is also essential to ensure the fair distribution of data points in clusters.

Let us consider $\textbf{X}$ as a binary cluster association (pilot assignment) matrix, where $x_{k,p}=1$ if \gls{ue} $k$ belongs to cluster (pilot) $p$, and $x_{k,p}=0$ otherwise.
So the repulsive clusters can be obtained by solving the following problem:
\begin{equation}
\begin{aligned}\label{eq:repOp}
    \max_{\textbf{X}} \quad & \sum_{p=1}^{\tau_p} \sum_{k=1}^{K-1}\sum_{k\prime=k+1}^{K} x_{k,p}x_{k',p}f_r(k, k') \\
    \textrm{s.t.} \quad & \sum_{p=1}^{\tau_p}x_{k,p}=1, \quad k \in \{1,...K\}\\
    \quad & \floor*{\frac{K}{\tau_p}} \leq \sum_{k=1}^K x_{k,p} \leq \floor*{\frac{K}{\tau_p}}+1, \quad p \in \{1,...\tau_p\}\\
    \quad & x_{k,p}\in\{0,1\}, k \in\{1,\ldots,K\}\,;
\end{aligned}
\end{equation}
where $f_r(k, k')$ is a customized function that measures the diversity/repulsion score for $k$ and $k'$ data points (\glspl{ue}).
The first constraint guarantees that each data point is assigned to one cluster and the second constrain forces the clusters to have similar size.
The second constraint is important because it keeps inter cluster similarity high.
This repulsion function can be a predefined static function, i.e., Euclidean distance, or can be parameterized and then learned by, e.g., neural networks.
The second approach is favorable as a sophisticated pilot assignment should consider not only the physical location of the \glspl{ue} but also other parameters like \gls{ap} locations and their density.

Repulsive clustering has already been considered in the literature under different names: anticlustering~\cite{papenberg2021using, brusco2020combining}, and maximally diverse grouping problem~\cite{schulz2021balanced}.
Typically this type of problem is NP-hard, but applying some relaxations can be solved by integer programming~\cite{grotschel1989cutting}.
Here, we present \Cref{alg:hrc}, a simple heuristic yet efficient algorithm to find a feasible (but not necessarily optimal) solution to the the repulsive clustering problem.
This algorithm first randomly assigns data points to different clusters and then iteratively swaps the \glspl{ue} among clusters as long as it improves the overall repulsion score.

 \begin{algorithm}
    \caption{A Heuristic Algorithm for Repulsive Clustering} 
    \label{alg:hrc}
    \textbf{Input:}
     Number of clusters (pilots) $\tau_p$, Set of \glspl{ue} $\mathcal{K}$\;\\
    \textbf{Output:}
    Pilot assignment vector $\textbf{p}$
    \begin{algorithmic}
            \State Randomly divide $\mathcal{K}$ UEs into $\tau_p$ equal-sized clusters $\mathcal{C}$,
            \While{Performance is improving}
                \For{$C1, C2 \in \mathcal{C}$}
                    \For{$u \in C1\ and\ w \in C2$}
                        \If{exchanging clusters of $u$ and $w$ increases the overall diversity measure as given by \eqref{eq:repOp}}
                            \State Swap the clusters of $u$ and $w$,
                        \EndIf
                    \EndFor
                \EndFor  
            \EndWhile
            \For{$p = 1:\tau_p$}
        \State Assign pilot $\phi_p$ to UEs in cluster $\textbf{p}$
    \EndFor
    \end{algorithmic}
\end{algorithm}

\section{Numerical results}\label{sec:result}
\subsection{Simulation setup}
Let us consider $M$ \glspl{ap} and $K$ \glspl{ue} that are independently and uniformly distributed in a $1\times 1$ km$^2$ square area.
We adopt the wrap-around technique to avoid boundary effects at the edge and simulate network behavior in an unlimited area.
The 3GPP Urban Microcell model~\cite{3gpp2010further} is used to compute the large-scale propagation conditions like path loss and shadow fading.
Noise power is calculated by $P_n=Bk_BT_0W$, where $B=20$ MHz is the bandwidth, $k_B=1.381\times 10^{-23}$ (Joule per Kelvin) denotes the Boltzmann constant, $T_0=290$ (Kelvin) is the noise temperature and $W=9$ represents the noise figure.
The transmission powers of the uplink pilot and the uplink data are set to $\rho_p = 100$ mW and $\rho_u = 100$ mW, respectively.
The channel estimation overhead has been taken into account for defining the per-user uplink throughput as $T_k^u = B\frac{1-\tau_p/\tau_c}{2}\log2(1+\operatorname{SINR}^u_k)$, where $\tau_c=200$ samples. The $1/2$ in the above equation is due to the co-existence of the uplink and downlink traffic.
We also employed max-min power control~\cite{ngo2017cell} to further improve the sum throughput.

In this paper we consider the Euclidean distance for the repulsion function as $f_r(U_k,U_{k'}) = \sqrt{\sum_{i=1}^{|F|} (U_k[i] - U_{k'}[i])^2}$,  where $F$ is the feature set (e.g. geographical coordinates) of the \glspl{ue}.
The definition and analysis of more sophisticated repulsive functions are left to future work. 

\subsection{Result and discussion}

\begin{figure}[t]
    \centering
    \includegraphics[width=\linewidth]{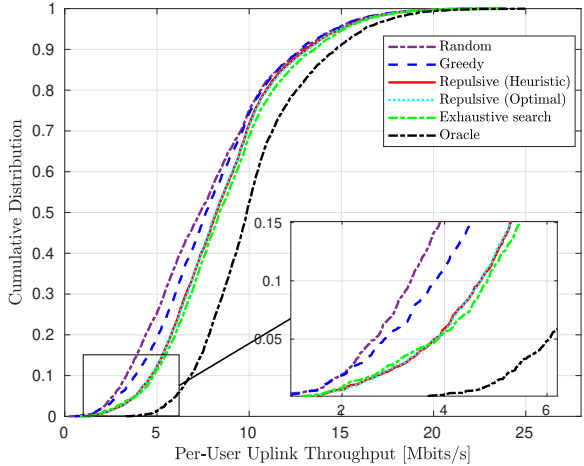}
    \caption{Cumulative distribution of the per-user uplink throughput for different pilot assignment strategies for a small-scale scenario, $M=50$, $K=12$ and $\tau_p=3$. }
    \label{fig:smallScenario}
\end{figure}

The \gls{cf} \gls{mmimo} systems aim to provide a uniform service to all the \glspl{ue} regardless of their physical location.
So, the per-user throughput is used to evaluate the performance of the pilot assignment algorithms.
The result is compared with the \textit{random} and \textit{greedy} pilot assignment from \cite{ngo2017cell} and \textit{Oracle} pilots assignment, where there is no pilot contamination.
The R-package introduced in~\cite{papenberg2021using} is used to optimally partition \glspl{ue} into diverse groups (optimal repulsive clustering).

As the time complexity of exhaustive search and optimal repulsive clustering exponentially grows by the number of \glspl{ue}, calculating their performance for large $M$s is not possible.
\Cref{fig:smallScenario} shows the \gls{cdf} of the per-user uplink throughput for a small-scale scenario, for the sake of comparison.
As seen in the figure, the method outperforms the \textit{random} and \textit{greedy} pilot assignments and basically achieves the same (optimal) performance of the exhaustive search, but with far less complexity.

\begin{figure}[t]
    \centering
\includegraphics[width=\linewidth]{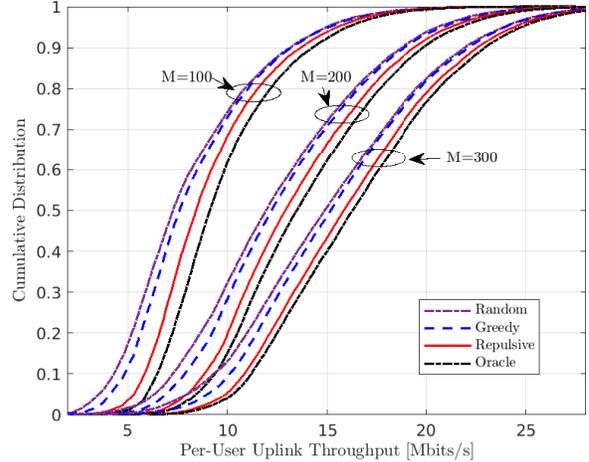}
    \caption{Cumulative distribution of the per-user uplink throughput for different pilot assignment strategies, $K=40$ and $\tau_p=10$. }
    \label{fig:cdf}
\end{figure}

\Cref{fig:cdf} shows the cumulative distribution of the per-user uplink throughput for different pilot assignment strategies for $M=\{100, 200, 300\}$.
The superiority of the proposed scheme against other approaches by a high margin is evident from the figure.
The decreasing gap between the repulsive and \textit{Oracle} pilot assignment by increasing the number of \glspl{ap} shows the robustness of our approach against density.

\begin{table}[]
    \centering
    \begin{tabular}{ccccc}
    \hline
    $M$ & Random & Greedy & Repulsive & Oracle \\ \hline
    100 & 3.5  &  4.1  &  5.3  &  5.9 \\\hline
    200 & 6.3  &  6.9  &  7.9  &  8.4 \\\hline
    300 & 7.9  &  8.9  &  9.9  & 10.3 \\\hline
    \end{tabular}
    \caption{95th percentile of the per-user uplink throughput [Mbits/s] for different numbers of APs, Here, $K=40$ and $\tau_p=10$.}
    \label{tbl:tbl1}
\end{table}
\Cref{tbl:tbl1} shows the 95th percentile of the per-user throughput extracted from \Cref{fig:cdf}.
Our method, for $M=100$, increases the 95th percentile of the per-user throughput by 1.75~Mbps (33\%) and 1.18~Mbps (22\%) in comparison to \textit{random} and \textit{greedy} assignments, respectively.
The improvements for $M=300$ are 1.93~Mbps (20\%) and 0.92~Mbps (9\%).
Compared to the situation without pilot contamination, our method successfully reaches 89\% to 96\% of the 95th percentile of the per-user throughput of \textit{Oracle} pilot assignment, which is a great success.

\begin{figure}[t]
    \centering
    \includegraphics[width=\linewidth]{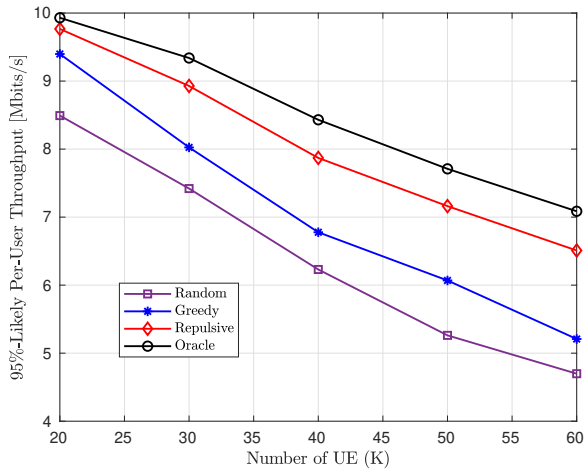}
    \caption{95th percentile of the per-user uplink throughput for different numbers of UE, $M=100$ and $\tau_p=10$.}
    \label{fig:likelyM}
\end{figure}

\Cref{fig:likelyM} illustrates the 95th percentile of the per-user uplink throughput of different pilot assignment schemes against the number of \glspl{ue}.
It can be seen from the figure that by increasing the number of \glspl{ue} in the network, the throughput for most of the \glspl{ue} decreases, but the reduction speed varies for different approaches.
The increasing gap between the proposed and other approaches shows the superiority of our system.
For example for $K=60$, our method improves the 95th percentile of the per-user throughput by 1.8~Mbps (38\%) and 1.3~Mbps (25\%), comparing to the \textit{random} and \textit{greedy} approaches, respectively.
Also, the gap between the proposed and the \textit{Oracle} pilot assignment scheme grows much slower than for the two other methods, which means that increasing $K$ does not heavily affect our system.

\begin{figure}[t]
    \centering
    \includegraphics[width=\linewidth]{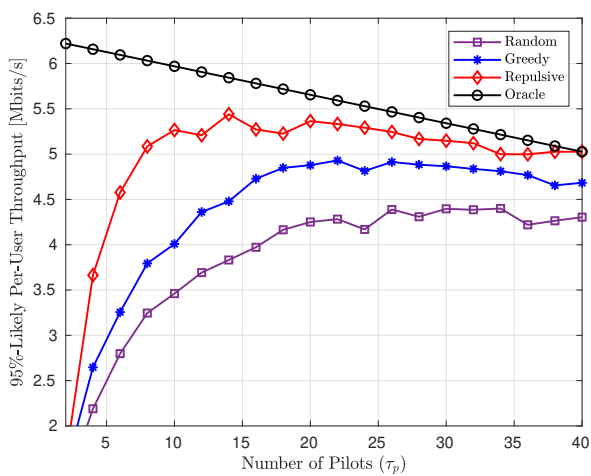}
    \caption{95th percentile of the per-user uplink throughput for different numbers of UE, $M=100$ and $K=40$.}
    \label{fig:likelyTau}
\end{figure}

The 95th percentile of the per-user uplink throughput against the number of pilots ($\tau_p$) for different pilot assignment schemes is presented in \Cref{fig:likelyTau}.
The proposed approach always performs better than the \textit{greedy} and \textit{random} pilot assignments.
As can be seen in the figure, increasing the number of pilots will improve the performance for the majority of the \glspl{ue} only up to a certain point, and reduces afterward.
This shows the necessity of finding the optimal number of pilots, a study which is outside the scope of this paper and is left for future research.

\section{Conclusion}\label{sec:conclusion}
In this paper, we proposed a repulsive clustering based pilot assignment for \gls{cf} \gls{mmimo} systems.
We formulated the pilot assignment as a maximally diverse clustering problem and solved it by a repulsive clustering scheme.
Numerical results show the effectiveness of the proposed scheme compared to the conventional random and greedy pilot assignment.
In future works, we will expand our approach by replacing the Euclidean distance with more sophisticated and parameterized repulsion functions, i.e., \glspl{dnn} that consider different networking factors such as \gls{ap} locations, and the density of \glspl{ue} and \glspl{ap}.
Another extension will consider pilot assignment jointly with pilot power control, which can further improve the channel estimation performance.
The scalability of different pilot assignment strategies is another factor that should be considered in future research.


\ifCLASSOPTIONcaptionsoff
  \newpage
\fi

\bibliographystyle{IEEEtran}
\bibliography{PARC_v02_arxiv.bib}



\end{document}